\documentstyle[12pt]{article}
\input math_macros
\author{Soo-Jong Rey\thanks{Yale-Brookhaven
SSC Fellow. Bitnet: Soo@Yalehep. This work was supported in part by
Texas National Research Laboratory Commission.}\\
\\
\it{Institute for Theoretical Physics}\\
\it {University of California, Santa Barbara CA 93106}\\
\it{\&}\\
\it{Center for Theoretical Physics, Sloane Laboratory}\\
 \it{Yale University, New Haven CT 06511 USA}}
\title{
{Spontaneous CP Violation}\\
{Small Instanton \& Invisible Axion}}
\date{September 1992}

\begin{document}

\maketitle
\begin{picture}(0,0)(0,0)
\put(300,320){NSF-ITP-92-107}
\put(310,300){YCTP-P20-92}
\end{picture}
\vspace{-5pt}

\begin{abstract}
I propose a variant invisible axion model of spontaneous
CP violation at the electroweak scale without
CP domain-wall and `strong CP' problems.
Both large sized QCD and small sized non-QCD instantons
break CP and Peccei-Quinn symmetries, and render
cosmologically harmful CP domain walls unstable.
The decaying epoch
depends on size of small instanton effects, and
is around $100 eV$ if the current neutron electric
dipole moment bound is maximally saturated.
The model satisfies constraints
from primordial D and He photo-dissociation and
black hole formation, while producing
cosmologically interesting size of gravitational
waves and galaxy-scale density perturbations.
\end{abstract}

\endtitlepage
\sl 1. Motivation: \rm \hfill\break
Origin of CP violations\cite{lincolnreview} still remains elusive.
Current experimental data from $K^o - \bar K^o$ system and
neutron electric dipole moment (NEDM):
$$
\epsilon \le 2.26 \cdot 10^{-3}; \hskip1cm |{\epsilon' \over \epsilon}| \le
1.45 \cdot 10^{-3};
\hskip1cm
\bar \theta_{\rm qcd} \le 10^{-13} ({m_N \over m_q}).
\eqno (1)
$$
indicate
\sl strong violation \rm of weak, flavor nondiagonal CP
(FNCP) symmetry and
\sl weak violation \rm of strong, flavor diagonal CP (FDCP) symmetry.

Complex Yukawa coupling constants or
complex Higgs coupling constants are known to provide sources of
explicit CP violations.
Alternatively, CP violations may arise spontaneously\cite{tdlee}
\cite{weinberg}
during the electroweak symmetry breaking, for example, as in
extended technicolor theories\cite{etc}. In addition, in
supersymmetric standard
models, spontaneous CP violation offers an attractive
solution\cite{susyscpv} of
suppressing other potentially large contributions to
NEDM than $\bar \theta_{\rm qcd}$.
It has been known, however, that this option is cosmologically
disastrous due to appearence of stable CP domain walls in the early
universe\cite {okun}. The walls quickly dominate
energy density of the universe, and lead too short the present age
to our Universe.

Recently, I proposed a particle physics solution to this
cosmological CP domain wall problem\cite{kraussrey},
utilizing nonperturbative QCD effect.
The idea was as follows. QCD instanton effect is practically
negligible around the electroweak scale at which CP is spontaneously broken.
However, the effect becomes pronounced as the QCD scale $\Lambda_{\rm qcd}
\sim 0.2 GeV$
is approached and induces operators
$$
{\cal L}_{nonp} = K_{\rm qcd} e^{i \bar \theta_{\rm qcd}}
(\bar U_L U_R \bar D_L D_R \bar S_L S_R)  + h.c.
\eqno (2)
$$
Under CP transformation,
$$\eqalign {
&\hskip1cm \bar \theta_{\rm qcd} \, \rightarrow \, \, -
\,\, \bar \theta_{\rm qcd} \cr
& (CP) {\cal L}_{nonp} (CP)^{-1} \ne {\cal L}_{nonp}.}
\eqno (3)
$$
Unless
$\bar \theta_{\rm qcd}$ is $0$ or $\pi$, CP, a discrete
global symmetry, is explicitly broken by the QCD instantons.
Being a nonperturbative effect, size of the intrinsic CP violation
is tiny;
$K_{\rm qcd} \sim \Lambda_{\rm qcd} m_u m_d m_s <\!\!< \, M_W^4$.
This lifts the vacuum degeneracy connecting two sides of the CP domain wall,
and leads to a subsequent wall annihilation driven by a false vacuum decay.

A drawback of this idea was the `strong-CP' problem
in assuming  $\bar \theta_{\rm qcd}$ nonzero but small
$\sim 10^{-9}$ to meet Eq.(1).
One may introduce Kim's\cite{ksvz} (KSVZ) or
Dine-Fischler-Srednicki\cite{dfsz} (DFSZ) invisible axion
as a viable solution to the `strong-CP' problem.
However, in the same paper\cite{kraussrey}, I also have shown
that spontaneous CP violation and the Peccei-Quinn (PQ) symmetry are
not compatible each other, at least, in the simplest two-Higgs model.
It was found that there always exists a nonanomalous discrete
symmetry, which is a simultaneous CP and finite PQ transformations:
$$
D: \,\,\,\,\, (PQ)_{2 \bar \theta} (CP) \, {\cal L} \, (CP)^{-1}
(PQ)^{-1}_{2 \bar \theta} = {\cal L}.
\eqno (4)
$$
Thus, there remains yet another cosmological domain wall problem
when the new discrete symmetry $D$ is spontaneous broken at the
electroweak scale.
Furthermore, especially in DFSZ axion model, I also observed
that the PQ symmetry drives the Higgs potential minima to a
CP-conserving point at least in the simplest two electroweak Higgs model.

In this paper, I propose a viable solution to these difficulties,
and make PQ symmetry compatible with spontaneous CP violation
at the electroweak scale. I achieve this by (1) extending
the electroweak Higgs sector and (2) taking into account of
additional nonperturbative effects from non-QCD sector, for
example, technicolor gauge group or semiclassical gravity.
Point (1) allows a nontrivial EW vacuum in which CP is
spontaneously broken, while point (2) lifts a vacuum degeneracy
connected by the discrete symmetry of Eq.(4). Furthermore, the
unstable but long-lived domain wall may serve an extra source of
long-wavelength gravitational waves and galaxy scale density perturbations.

\sl 2. Model:\rm \hfill\break
I first state criteria what a viable model is required to meet:

(1) \sl intrinsically broken strong CP and an invisible axion solution to
the `strong CP' problem;

(2) spontaneously broken weak CP during the electroweak phase transition. \rm
\vskip0.5cm

(Our viewpoint is different from Nelson-Barr\cite{nelsonbarr} in that
flavor-nondiagonal CP begins to appear at the electroweak scale,
and that `strong-CP' problem is solved by an invisible axion.)

I introduce a singlet Higgs field $\Sigma$ and add one more
to weak doublet Higgs fields $H_i$ in the standard axion models.
Thus, the standard KSVZ model is now extended to Lee's two-Higgs
doublet model\cite{tdlee}, while the standard DFSZ model to Weinberg's
three-Higgs doublet model\cite{weinberg}. We do not, however, introduce
Glashow-Weinberg (GW) criterion\cite{glashowweinberg}
of natural flavor conservation (NFC);
approximate flavor symmetry for small Yukawa coupling constants is
already technically natural\cite{hall}.
The GW criterion usually imposes an additional discrete symmetry to
the Yukawa and Higgs sector, which is
spontaneously broken at the electroweak symmetry breaking.
Cosmologically, this causes yet another domain wall problem,
which I wanted to avoid.

$\Sigma$ field acquires a vacuum expectation value (VEV)
$<\!\! \Sigma \!\!> \, \approx f_a \sim 10^{12} \, GeV$, and
breaks PQ symmetry spontaneously. Radiative corrections
do respect PQ and CP symmetries as exact symmetries.
Thus,
effective electroweak Higgs potential near the weak scale
consists of all renormalizable  gauge, PQ and CP invariant operators
among Higgs doublets $H_i$ and axion $a \sim arg \Sigma$. Schematically,
it reads
$$  \eqalign{
V_{\rm Higgs} [H, \Sigma] = & \sum_{ij} M_{ij}(\Sigma) H_i^\dagger H_j \cr
+ & \sum_{ijkl} \lambda_{ijkl} (\Sigma) (H^\dagger_i H_j) (H^\dagger_k H_l)
+ \sum_{ijkl} \tilde \lambda_{ijkl} (\Sigma) (H^\dagger_i \vec \tau H_j)
(H^\dagger_k \vec \tau H_l).}
\eqno (5)
$$
Hermiticity and functional dependence of
$M^2, \, \lambda, \, \tilde \lambda$ to the axion field $\Sigma$
is fixed by CP and PQ invariances of
the low-energy Higgs potential $V_{\rm Higgs}$.
Low-energy Yukawa couplings are similarly constrained.
Minimization of the above electroweak Higgs potential was analyzed
previously\cite{weinberg}\cite{kraussrey}.
For both Lee-KSVZ and Weinberg-DFSZ models,
it can be shown that there does exist a region of parameter space in
which the CP and gauge symmetries are spontaneously broken:
$<\!\! \phi_a\!\!> = {v_a \over \sqrt 2} e^{i \theta_a}$ with
$\theta_a - \theta_b \ne n \pi$.
(Related multi-Higgs models were studied previously\cite{geng}.)

In addition, QCD and non-QCD instantons generate additional terms
to the electroweak Higgs potential.
An example of non-QCD nonperturbative effect is the gravitational
instanton.
The simplest gravitational instanton with an asymptotically flat
spacetime is the Eguchi-Hanson metric\cite{eh}.
The EH instanton does not break the PQ symmetry. On the other hand,
a charged EH instanton can do.
An EH instanton of size $\rho$ with a self-dual
electromagnetic field (EM instanton) is known\cite{eh}:
$$
g^{\mu \nu} = \delta^{\mu \nu} -
{\rho^4 \over x^4} {x^\mu x^\nu \over x^2}
+ {\rho^4 \over x^4 -\rho^4} {\tilde x^\mu \tilde x^\nu \over x^2};
\hskip 1cm A_\mu = 2P {\rho^2 \tilde x^\mu \over x^4}
\eqno (6)
$$
in which $x^2 \equiv x^\mu x^\mu \ge \rho^2$ and $\tilde x^\mu = (y, -x, t,
-z)$ is identified with $x^\mu$ because of ${\bf R}\! {\rm P}_3$ spacetime
topology. The instanton EM charge $P \in 3 {\bf Z}$ to give a compatible
spin structure for the standard model chiral fermions.
Instanton action is found $S_{\rm EM} = {\pi P^2 \over \alpha_e(\rho)}
>\!\!> 1$,
hence, semiclassical approximation is valid.
In the QCD and EM instantons backbround, the PQ current is anomalous:
$$
\nabla^\mu J_{\mu \rm pq} = \sum_{\rm ferm} Q_{\rm pq}
( {2 Q_{\rm em}^2 \over 16 \pi^2} F_{\mu \nu} \tilde
F_{\mu \nu} + {2 C_2 ({\bf R}) \over 8 \pi^2 }
{\rm Tr} {\bf G}_{\mu \nu} \tilde {\bf G}_{\mu \nu})
\eqno (7)
$$
in which the sum is over all fermions with nontrivial PQ charge.
In the dilute instanton approximation, the instantons
induce multi-fermion operators similar to Eq.(2).
They constitute the \sl only \rm sources of intrinsic PQ and CP symmetry
breaking (potentially dangerous irrelevant operators).
Tying up fermion zero modes with PQ and CP conserving Yukawa
and Higgs self-interaction vertices, a low-energy
Higgs-axion potential is found;
$$
V_{\rm inst} = \int_{f_a^{-1}}^{\Lambda_{\rm qcd}^{-1}} d \mu [\rho]\,
[K_{\rm qcd} (\rho) e^{i \bar \theta_{\rm qcd}}
\Sigma^m + K_{\rm em} (\rho) e^{i \bar \theta_{\rm em}}
\Sigma^n + h.c.].
\eqno (8)
$$
Magnitude of EM instanton induced axion potential $K_{\rm em}$ was
estimated in detail\cite{holmankephartrey} to give
$ K_{\rm em} \sim 10^{-9} K_{\rm qcd}$ in which $K_{\rm qcd}
\sim 10^{-6} \, GeV^4$.
Thus, modifications in the above
electroweak Higgs potential minimization due to QCD and EM instantons
are totally negligible.
The CP-odd
coupling constants $\bar \theta$'s do receive finite renormalizations
across each fermion mass thresholds and from the spontaneous
CP violating phases $\theta_{ab}$.
Instanton multiplicity $m, \, n \in {\bf Z}_+$ does depend on
details of axion model. Since each term in Eq.(8) comes from
different instanton effect, I expect $m \ne n$ in general.
QCD and
EM instantons give:
$$
m \equiv {\rm Tr} 2 C_2 (R); \hskip1cm n \equiv {\rm Tr} 18 Q_{\rm em}^2
\eqno (9)
$$
in which trace is over all fermions with nonzero PQ charge assignment.

Each of the operators appearing in Eq.(8) breaks CP and PQ symmetries
separately. If either $K_{\rm qcd}$ or $K_{\rm em}$ were absent,
the combined Higgs potential $V_{\rm Higgs}  + V_{\rm inst}$
is manifestly invariant under simultaneous transformations of
CP: $\bar \theta \rightarrow -\bar \theta$ and a finite PQ: $ arg \Sigma
\rightarrow arg \Sigma - 2/m \, ({\rm or} \,\, 2/n) \cdot \bar \theta$.
This is the new discrete symmetry $D$
alluded earler in Eq.(4), and is the origin of another cosmological
domain wall problem.

Important point is that this ceases to be the case if \sl both \rm
$K$'s are present.
CP transformation changes:
$$
\bar \theta_{\rm qcd}, \,\, \bar \theta_{\rm em} \rightarrow -
\bar \theta_{\rm qcd},
\,\, - \bar \theta_{\rm em}.
\eqno (10)
$$
Thus, \sl no \rm subsequent finite PQ transformation
$$
arg \Sigma \rightarrow \Sigma - \alpha
\eqno (11)
$$
can rotate both $\bar \theta$'s back to the original values
\sl simultaneously \rm. Since PQ symmetry is the only other available
symmetry to rotate phases in $V_{\rm inst}$, I draw a conclusion that
there simply does not exist any residual,
nonanomalous discrete symmetry such as $D$ in Eq.(4).

The size of additional instanton induced operator, $K_{\rm em}$
is further constrained. Minimizing the axion potential Eq.(8),
I find
$$
m \, <\!a\!> + \bar \theta_{\rm qcd} \approx { K_{\rm em} \over
K_{\rm qcd}} \cdot ({n^2 \over m^2} \bar \theta_{\rm qcd} -
{n \over m} \bar \theta_{\rm
em}).
\eqno (12)
$$
Taking
$|\bar \theta_{\rm qcd} - \bar \theta_{\rm em}| \sim {\cal O}(\pi)$,
this implies
$$
K_{\rm em}  \le 10 ^{-9} K_{\rm qcd}
\eqno (13)
$$
to be consistent with NEDM measurement Eq.(1).
As obvious as it may sound, however, nonzero value of
$\bar \theta \equiv m \! <\! a \!> + \bar \theta_{\rm qcd}
(1 \, GeV) \sim 10^{-9}$
makes the CP domain wall unstable as well!

Energy density difference between the otherwise degenerate vacua connected
by $D$ symmetry
is estimated from the potential Eq.(8):
$$
\Delta {\cal E}_{vac} \approx K_{\rm em}
\sin 2 ({n \over
m} \bar \theta_{\rm qcd} - \bar \theta_{\rm em})
\sin (n \! <\! a\!> + \bar \theta_{\rm em}).
\eqno (14)
$$
In case Eqs. (12) and (13) are maximally saturated,
$$
\Delta {\cal E}_{vac} \approx 10^{-9} K_{\rm em} \approx (100 \, eV)^4.
\eqno (15)
$$
It is utterly small compared to the QCD scale, yet large enough
to be cosmologically significant as shall be discussed below.

\sl 3.  Cosmological Evolution of Unstable Domain Wall: \rm \hfill\break
Cosmological evolution of the CP domain wall was studied in my
previous
work\cite{kraussrey}. At the epoch PQ symmetry is spontaneously broken,
the QCD and EM instanton induced potential Eq.(6) is
 negligible.
This remains true even until an electroweak transition
temperature $T_{EW}$ is reached and
the CP domain walls are formed.
As the temperature goes down further,
energy density of the practically stable domain walls
starts to grow faster than
that of the background radiation.
Due to background plasma viscosity, the domain wall is straightened
out only to a scale
$R_c (T) \sim \sqrt {M_w^3 M_{\rm pl}} /T^3$, which is a fraction
$F \equiv R_c (T) H(T)$ of the horizon at a given temperature $T$.
Using this, epoch that energy density of the wall begins dominate over
that of
radiation is found as\cite{kraussrey}:
$$
\rho_{\rm wall} \approx \rho_{rad} \hskip0.5cm \rightarrow \hskip0.5cm
T_c \approx 10^{-8} {1 \over \sqrt \lambda} T_{EW}
\eqno (16)
$$
most conservatively, in which
$\lambda$ denotes a typical value of
Higgs quartic couplings $ \lambda_{ijkl} $ and $\tilde \lambda_{ijkl}$.
With $\lambda \sim 10^{-4}$,
the domain wall starts to dominate energy density of the universe
only below $T_c \approx 300 eV$.
In particular, this implies that primordial nucleosynthesis (PNS) at the
MeV scale is not disturbed by the presence of CP domain walls, as their
energy density is completely negligible at the PNS epoch.

The otherwise stable CP domain wall begins to feel the
energy density difference in Eq.(8) as the pressure provided by the vacuum
energy density difference begins dominate over the viscosity
$\Delta {\cal E}_{vac} \ge T^4$.
This occurs when
$$
T^* \approx 10^{-6} \Lambda_{\rm qcd} \approx 100 \, eV.
\eqno (17)
$$
The wall then moves quickly to the speed of light via false vacuum
decay. As the wall reaches the speed of light,
regions of false vacuum are quickly driven away.

That $T^* \le {1 \over 3} T_c$ implies a brief period
of domain-wall dominated universe.
This period is prolonged if the EM instanton size $K_{\rm em}$
is smaller than the estimate I used; $\sim 10^{-9} K_{\rm qcd}$.
This transient, domain-wall dominated epoch near $T^*$ is after
the present day galaxy
sized regions first came inside the horizon around $T \sim 1 KeV$.
Relativistic motion of the walls tend to wash out primordial density
fluctuations smaller than the horizon scale, i.e., the galaxy scale.
Instead, randomly located walls generate a Poisson
distribution of density perturbations at scales smaller than the
present galaxy size. Magnitude of this perturbation is therefore
expected to be $\delta \rho /\rho \sim (R_c(T) H)^{1/2} \approx F^{1/2}$
{}.
Furthermore, as the domain wall dominated epoch gives rise to a modest
power-law
inflation $H (t) = 2/ t$,
it is expected that the density perturbation of galaxy size
is further damped, and may provide a `biasing' mechanism to the standard
galaxy formation.

Violent wall collision will produce both strong metric perturbation
which in turn forms
gravitational waves (GW)
and primordial
black holes (PBH).
Since the walls move quickly with a relativistic velocity,
time scale of collision is expected to be $\sim R_c (T^*)$. Thus
radiated gravitational wave energy $E_{\rm gw}$ and
wavelength $\lambda (T^*)$
with a quadrupole
$Q$ is estimated:
$$
E_{\rm gw} = {1 \over M_{\rm pl}^2} ({d^3 Q \over dt^3})^2
 \approx {1 \over M_{\rm pl}^2} {E_v^2 \over R_c};
\hskip0.7cm \lambda(T^*) \sim R_c (T^*)
\eqno (18)
$$
in which $E_v = {4 \pi \over 3} R_c^3 \Delta {\cal E}_{\rm vac}$.
Therefore, we find $\Delta E_{\rm gw} / E_v \approx F^2$
and is about $15 - 25 \%$ at $T = T^*$.
Thus, a significant fraction of the false vacuum energy goes
into the gravitational wave radiation\cite{turnerwilczek}.
Using the fact that the gravitational wave vector $2 \pi / \lambda$ and
amplitude $h(\lambda)$ both
scales as $1 \over R(T)$, I finally
estimate them at present time as:
$$
h (\lambda) \approx 10^{-8};\hskip1cm
\lambda \approx 10^{23} (cm).
\eqno (19)
$$
Both the amplitude and frequency fall short an order of
magnitude of interesting
ranges that millisecond pulsar and large scale cosmic microwave
background
radiation experiments\cite{stinebring} may be able to detect!
Uncertainties in the above rough estimate should be reflected in
Eq.(19); a better quantitative estimate might be able to
put the present model against upcoming cosmological experimental tests.

Similarly, PBH's may form as the wall collide.
With the wall-induced density fluctuation $\delta \rho / \rho$
estimated to $\sim F^{1/2} \sim 50-60\%$ I expect the PBH formation
inefficient even at the locus of wall collisions.
Typical mass of PBH's is estimated heavy enough
$\sim \epsilon T^{*4} \cdot (R_c^2
/H) \approx \epsilon F^2 (T^{*4} /H^3)$ that its Hawking radiation
is completey negligible. Here, $\epsilon$ denotes
a geometrical factor for the locus of wall collisions at which the
PBH's are formed, and is expected to be $ \epsilon <\!\!< 1$.
Maximal fraction of the PBH formation energy
$\Delta E_{\rm pbh}$ is constrained,
in case the practically stable PBH's dominate the matter of
the present universe,
by
$\Delta E_{\rm pbh} /E_v \le (13.6 eV / T^* )\sim 13 \%$.
This bound is viewed large enough to accomodate even the most efficient
PBH formation possible.
Correspondingly, PBH produced gravitational waves are negligible
compared to the one generated by wall collision alluded above.

The wall collision also produces ${\cal O} ({1 \over \lambda})$
energetic Higgs $H^*$ and subsequently
cascading lighter particles including photons $\gamma^*$,
to which PNS puts a strong constraint.
The energetic photons may
photo-dissociate and change D and He abundances from the successfully
predicted values of the standard PNS model.
However, the energetic particles scatter predominantly with the
$10^9$ more abundant, cosmic background photons $\gamma_{\rm cb}$
than D's and He's.
This will continue until $T \sim 10^3 \, eV$\cite{sarker} at which number of
the
high-energy tail
$\gamma_{\rm cb}$'s
is comparable to the number of D and He's, and threshold of
$H^* + \gamma_{\rm cb}$ or $\gamma^* + \gamma_{\rm cb}
\rightarrow l \bar l$ is reached.
This is precisely the scale that CP domain walls completes decaying.
Thus, PNS constraint of D and He abundances
is seen marginally compatible with the present model.

I thank L.M. Krauss, L. McLerran and M. Srednicki for useful discussions.
I also acknowledge hospitality of the Institute for Theoretical
Physics at Santa Barbara where this research was finished.
This work was supported in part by the National Science Foundation
under grant No. PHY89-04035, and by DOE and TNRLC.

\bibliographystyle {unsrt}

\begin{thebibliography}{99}

\bibitem {lincolnreview} For an comprehensive collection of
original papers, see L. Wolfenstein,
\sl CP Violation, \rm North-Holland Pub. Co. (1989).

\bibitem {tdlee} T.D. Lee, Phys. Rev. \bf D8 \rm (1973)
1326; Phys. Rep. \bf 9 \rm (1974) 143, 176; see also,
J. Liu and L. Wolfenstein, Nucl. Phys. \bf B289 \rm (1987) 1.

\bibitem {weinberg} S. Weinberg, Phys. Rev. Lett. \bf 37
\rm (1976) 657; see also,
G. Branco, Phys. Rev. \bf D22 \rm (1980) 2901.

\bibitem {etc} E. Eichten, K. Lane and J.P. Preskill,
Phys. Rev. Lett. \bf 45 \rm (1980) 225.

\bibitem {susyscpv} N. Maekawa, Phys. Lett. \bf 282B \rm (1992) 387;
A. Pomarol, \sl Spontanoeus CP Violation in Supersymmetric Theories,
\rm SCIPP-92/30 preprint (1992).

\bibitem {okun} Ya.B. Zel'dovich, I.B. Kobzarev and
L. Okun, Phys. Lett. \bf B50 \rm (1974) 340.

\bibitem {kraussrey} L.M. Krauss and S.-J. Rey,
Phys. Rev. Lett. (1992).

\bibitem {ksvz} J.E. Kim, Phys. Rev. Lett. \bf 43 \rm (1979)
103; M.A. Shifman, V.I. Vainshtein and V.I. Zakharov, Nucl.
Phys. \bf B166 \rm (1980) 493.

\bibitem {dfsz} M. Dine, W. Fischler and M. Srednicki,
Phys. Lett. \bf 109B \rm (1981) 199; A.P. Zhitniskii, Sov.
J. Nucl. Phys. \bf 31 \rm (1980) 260.

\bibitem {nelsonbarr} A. Nelson, Phys. Lett. \bf 136B \rm
(1984) 387; S.M. Barr and A. Zee, Phys. Rev. Lett. \bf 55
\rm (1985) 2253; S.M. Barr, Phys. Rev. \bf D34 \rm (1986) 165.

\bibitem {glashowweinberg} S.L. Glashow and S. Weinberg,
Phys. Rev. \bf D15 \rm (1977) 1958.

\bibitem {hall} A. Antaramian, L.J. Hall and A. Rasin, \sl
Flavor Changing Interactions Mediated by Scalars at the Weak Scale,
\rm LBL-32345 preprint (1992).

\bibitem {geng} A.R. Zhitniskii, Yad. Fiz. \bf 31 \rm (1980) 1024 [Sov. J.
Nucl. Phys. \bf 31 \rm (1980) 529];
C.Q. Geng, X.D. Jiang and J.N. Ng, Phys. Rev. \bf D38 \rm (1988) 1628.

\bibitem {eh} T. Eguchi and A. Hanson, Phys. Lett. \bf 74B \rm (1978)
430:  Ann. Phys. (NY) \bf 120 \rm (1979) 82; G. `t Hooft, Nucl. Phys.
\bf B315 \rm (1989) 517; see also, S.-J. Rey, \sl Gravitational
Superstring Instantons and Solitons, \rm NSF-ITP-92-106, YCTP-P18-92 preprint
(1992).

\bibitem {holmankephartrey} R. Holman, T.W. Kephart and S.-J. Rey,
\sl Semiclassical Gravity and Invisible Axions, \rm NSF-ITP-92-85,
YCTP-P17-92 preprint (1992).

\bibitem {turnerwilczek} M.S. Turner and F. Wilczek, Phys. Rev. Lett.
\bf 65 \rm (1990) 3080.

\bibitem {stinebring} D.R. Stinebring \etal, Phys. Rev. Lett. \bf 65 \rm
(1990) 285.

\bibitem {sarker}
D. Lindley, Mon. Not. R. Astron. Soc. \bf 188 \rm (1979) 15P; \bf 193
\rm (1980) 593; L.M. Krauss, Phys. Rev. Lett. \bf 53 \rm (1984) 1976;
S. Sarker and A.M. Cooper, Phys. Lett. \bf 148B \rm (1984) 347.
\end{thebibliography}

\end{document}